\documentclass[twocolumn, prd, amssymb]{revtex4}
\usepackage{epsfig}
\def\IZ{\relax\ifmmode\mathchoice
{\hbox{\cmss Z\kern-.4em Z}}{\hbox{\cmss Z\kern-.4em Z}}
{\lower.9pt\hbox{\cmsss Z\kern-.4em Z}} {\lower1.2pt\hbox{\cmsss
Z\kern-.4em Z}}\else{\cmss Z\kern-.4em }\fi}
\def\IC{\relax\hbox{$\inbar\kern-.3em{\rm C}$}}
\def\IR{\relax{\rm I\kern-.18em R}}
\def\1{\relax 1 { \rm \kern-.35em I}}

\def\p{\prime}
\newcommand {\nn} {\nonumber}
\newcommand{\be}{\begin{equation}}
\newcommand{\ee}{\end{equation}}
\newcommand{\bea}{\begin{eqnarray}}
\newcommand{\eea}{\end{eqnarray}}
\newcommand{\JHEP}{J. High Energy Physics}

\def\ajou#1&#2(#3){\ \sl#1\bf#2\rm(19#3)}

\def\TrH#1{ {\raise -.5em
                      \hbox{$\buildrel {\textstyle  {\rm Tr } }\over
{\scriptscriptstyle \CH _ {#1}}$}~}}

\begin{document}
\preprint{NTUTH-02/S015} \preprint{CERN/TH/02-xxx}
\preprint{hep-th/0212092}
\title{Black hole entropy in string-generated gravity models}
\author{Ishwaree P. Neupane}
\affiliation{Department of Physics, National Taiwan University,
Taipei 106, Taiwan, R.O.C }

\date{December 30, 2002}

\begin{abstract}
The Euclidean action and entropy are computed in string-generated
gravity models with quadratic curvatures, and used to argue that a
negative mass extremal metric is the background for hyperbolic
($k=-1$) black hole spacetimes, $k$ being the curvature constant
of the event horizon. The entropy associated with a black hole is
always positive for $k=\{0,~1\}$ family. The positivity of energy
condition also ensures that the $k=-1$ (extremal) entropy is
non-negative.

\end{abstract}


\maketitle

The area--entropy law~\cite{Bekenstein73a} (in Planck units) $
{\cal S}=\frac{A_H}{4G}$ (where $A_H$ is the area of the event
horizon of the black hole and $G$ is the Newton constant) is one
of most celebrated results in general relativity. It is
known~\cite{Myers88a} that the black hole entropy is not simply
given by one-quarter the area, particularly, if one allows higher
curvature corrections to the Einstein action, such as  \bea
\label{action2} I&=&\frac{1}{16\pi G}\int
d^{n+1}x\,\sqrt{-g}\,\left(R-2\Lambda\right) +\alpha_1\int
d^{n+1}x \nn \\
&\times &
\sqrt{-g}\left(R_{\mu\nu\lambda\rho}R^{\mu\nu\lambda\rho}+a
R_{\mu\nu}R^{\mu\nu}+b R^2\right)+\cdots\,. \eea There are some
known reasons to explore black holes in such generalized gravity
models. The Gauss-Bonnet term obtained by setting $a=-\,4,~b=1$,
originally motivated by string theory, produces the most general
Lagrangian retaining only second-order field equations, and admits
exact spherically symmetric solutions in dimensions $n+1>
4$~\cite{Deser85a}. The action~(\ref{action2}) with $a=b=0$, $n=4$
corresponds to an effective $AdS_5$ (bulk) action, deduced from a
heterotic string via heterotic-type I duality~\cite{Blau99a}, \be
\label{5dAdS} I=\frac{N^2}{4\pi^2\,l^3}\int d^5x
\sqrt{-g}\left[\left(R-2\Lambda\right) + \frac{l^2}{16N}\,
R_{\mu\nu\lambda\sigma}R^{\mu\nu\lambda\sigma}\right]\,, \ee
where, using AdS/CFT duality~\cite{Maldacena97a}, the coefficient
of $(\mbox{Riemann})^2$ term is fixed as $32\pi
G\,\alpha_1=l^2/8N\equiv \varepsilon$.

One can evaluate leading order corrections to the black hole
entropy by finding exact solutions of Einstein equations
supplemented by higher curvature (HC) terms, such as a
Gauss-Bonnet term or quadratic interactions without a
$(\mbox{Riemann})^2$ term, or by treating HC terms as perturbation
about the Einstein gravity. The first approach allows one to study
global properties of the solutions with an asymptotically
(anti-)de Sitter branch~\cite{Deser85a,Myers88a,Nojiri01c,IPN02a}.
In this context, a question may be raised as to whether higher
derivative gravities can have negative entropy
~\cite{Nojiri01c,Odint02a}, in particular, when the curvature
length of AdS geometry itself is in the order of HC couplings. In
order to address this issue and gain some insight into the
problem, it is essential to calculate the total energy. In doing
so, we find that the requirement of positivity of energy ensures
the positivity of (extremal) black hole entropy.

In this paper, we also answer to the important question of what is
the correct ground state to use in hyperbolic anti-de Sitter
spacetimes. We reiterate the earlier assertions made by
Vanzo~\cite{Vanzo} and Birmingham~\cite{Birmingham98a} (see
also~\cite{Emparan99a} for a discussion in the context of the
counterterm substraction method) that a negative mass extremal
metric is the background for hyperbolic black
holes~\cite{Mann96a}.

The action~(\ref{action2}) with $a=-\,4$, $b=1$, admits the exact
black hole solution~\cite{Cai01a,IPN02a,Nojiri01c} \bea
\label{generalsol} ds^2&=&-f(r)\, dt^2+\frac{dr^2}{f(r)}+r^2
\sum_{i=1}^{n-1} h_{ij}
dx^i\,dx^j  \\
f(r)&=& k+\frac{r^2}{2\alpha}\mp
\frac{r^2}{2\alpha}\sqrt{1+\frac{8\alpha\Lambda}{n(n-1)}
+\frac{4\alpha\,\mu}{r^n}} \,.\label{solution}\eea where
$\alpha\equiv 16\pi G (n-2)(n-3)\alpha_1 $, $\mu$ is a mass
parameter, and $h_{ij}$ is the metric of an $(n-1)$-dimensional
maximally symmetric space ${\cal M}_k^{n-1}$ with curvature
$k=0,\,\pm 1$. For a symmetric space $R_{\mu\nu\lambda\rho}=
-\,(g_{\mu\lambda}g_{\nu\rho}-g_{\mu\rho}g_{\nu\lambda})/\ell^2$,
the cosmological term is fixed $\Lambda=-\,n(n-1)/2l^2$, where
$l=\ell/\sqrt{1-\alpha/\ell^2}$ is the (effective) curvature
radius of AdS bulk geometry. One also identifies the imaginary
time of the solution with a period $\beta=4\pi/f^\p(r_+)$, namely
\be \label{HawkingT} {\beta}=\frac{4\pi\, r_+\, l^2\,
\left(r_+^2+2\alpha\,k\right)} {n\,r_+^4+(n-2)\,k\, r_+^2\,l^2
+(n-4)\alpha\,k^2\,l^2}\,,\ee where $r_+$ is the largest positive
root of $f(r)$ (c.f., negative root in~(\ref{solution})), and
$1/\beta=T$ is the Hawking temperature of a black hole.

The extremal black-holes are defined to have zero temperature,
which require a vanishing denominator in~(\ref{HawkingT}).
Therefore, for $n=4$, there is an extremal $k=-1$ solution, with a
degenerate horizon at $r_+=r_e$, satisfying \be
r_{e}^2=\frac{l^2}{2}\,, \quad
\mu_{e}=-\frac{l^2}{4}\left(1-\frac{4\alpha}{l^2}\right)\,,
\label{4dextr}\ee and $\alpha<\frac{l^2}{4}$. Here we need to be
more precise. The above solutions are extremal ones only if
$\alpha<\frac{l^2}{4}$ holds. Because, in particular, for the
coupling $\alpha=l^2/4$, one obtains $\beta=\frac{8\pi r_+
l^2}{2n\, r_+^2+(n-4)k\,l^2}$, and hence the Hawking temperature
is finite, namely, $T=r_+/\pi l^2$, when $n=4$, independent of the
curvature $k$ of the horizon. That is to say, the extremal Hawking
temperature can be zero only for the coupling
$\alpha<\frac{l^2}{4}$~\cite{Cai01a}. To present a better picture,
we need to consider $n>4$ case. With $k=-1$, the $T=0$
($\beta=\infty$) condition yields \bea r_{c,\,
e}^2&=&\left(\frac{n-2}{2n}\right)l^2 \left(1\mp
\sqrt{1-\frac{4n(n-4)}{(n-2)^2}
\,\frac{\alpha}{l^2}}\right)\label{rextr}\\
\mu_{c,\,e}&=&\left(\frac{2 r_{c,\,e}^{n-2}}{n-4}\right)
\left[\frac{2}{n}\pm
\sqrt{\left(\frac{n-2}{n}\right)^2-\frac{4(n-4)}{n}\,
\frac{\alpha}{l^2}}\right].\label{mextr} \eea For $\alpha=0$, the
critical horizon $r_c~(<r_e)$ given by negative root
of~(\ref{rextr}) coincides with the singularity at $r=0$, so the
space-time region $r<r_{e}$ has an internal infinity. With
$\alpha>0$, we can have non-degenerate horizons with hyperbolic
geometry. Moreover, with $\alpha=l^2/4$, one has
$r_+=r_c=l\,\sqrt{(n-4)/2n}$ and hence
$T=\frac{n(r_+^2-r_c^2)}{4\pi r_+ l^2}=0$, which is of course not
a massless (BPS) state, since $\mu_c>0$. This corresponds to a
particular solution studied in~\cite{Zanelli00a}, where the
coupling $\alpha^\p$ is fixed in the starting action using
$8\alpha\Lambda+n(n-1)=0$. Notice that, for $\alpha=l^2/4$,
$\mu_e=0$ at $r_+=r_e=l/\sqrt{2}$ but $T\neq 0$. A clear message
is that only for $\mu_e<0$ (or $\mu_c>0$) background one can
consistently set $T=0$. The possible backgrounds are \bea \bullet~
n=4 &:& \mu_e=0\,,~T=\frac{1}{\sqrt{2}\,\pi l}~\mbox{or}~(ii)~
\mu_{e}< 0\,,~T=0 \,, \nn \\
\bullet~ n> 4 &:& \mu_e=0,\,~T>0; \mu_{c}>0,~ T=0; \mu_e<0,~T=0
\,.\nn \eea It would be natural to call "ground state" the state
with zero temperature. We find that only a negative mass extremal
state can be stable under gravitational (tensor) perturbations. So
a massless state may not be the ground state for the $k=-1$
horizon, as expected in~\cite{Vanzo,Birmingham98a,Surya01a}.

The on-shell Gauss-Bonnet gravity action reads \be
I=\frac{1}{16\pi G_{n+1}}\int
d^{n+1}x\,\sqrt{-g}\left(-\frac{2\,R}{n-3}
+\frac{8\Lambda}{n-3}\right)\,. \ee It is known that the AdS
space~\cite{Witten98a} and the Horowitz-Myers
soliton~\cite{Horowitz98a} are the appropriate backgrounds,
respectively, for spherical ($k=1$) and toroidal ($k=0$) horizons.
For $k=0$, a zero mass ground state is still legitimate, and is an
acceptable background~\cite{Vanzo,Birmingham98a}. For $k=-1$, by
matching the asymptotic geometries between extremal and
asymptotically locally AdS metrics, one subtracts a non-zero mass
extremal background~\cite{Birmingham98a}, restricting attention to
the region $r\geq r_e$ for the background and $r\geq r_+$ for the
black hole. The Euclideanized action, valid for $k=0,~\pm 1$, is
thus evaluated to be \bea \label{Eaction} \widehat{I} &=& -
\frac{(n-1) V_{n-1}\,r_+^{n-4}\,\beta}{16\pi G_{n+1}\,(n-3)}
\left[\left(k\,r_+^2-\alpha\,k^2\right)
+\frac{3\,r_+^4}{l^2}\right]\nn \\
&{}& +\frac{V_{n-1}r_+^{n-1}}{2(n-3)G_{n+1}}
-\frac{(n-1)V_{n-1}\,\beta} {16\pi G_{n+1}}\,\mu_{e}\,, \eea where
$V_{n-1}=\int d^{n-1}x\,\sqrt{h}$.
One reads off the free energy from $F=\hat{I}/\beta$. When
$\alpha=0,~k=-1$, there is no phase transition since the black
hole dominates over $\mu_{e}$ background for all temperatures.
Typically, a massless state at $\alpha=l^2/4>0$ has an initial
positive free energy in $n=4$ but zero free energy in $n=6$, so,
for $\alpha>0$ solutions, the behavior of Hawking-Page phase
transition could depend on spacetime dimensions, unlike in the
$\alpha=0$ case ~\cite{Hawking83a,Witten98a}.

Since $\mu_{e}$ is temperature (or horizon $r_+$) independent, the
black hole entropy takes a remarkably simple form \be
\label{entropyGB} {\cal S}=\beta^2\,\frac{\partial F}{\partial
\beta}=\frac{A_H}{4G_{n+1}}
\left[1+\frac{(n-1)}{(n-3)}\,\frac{2\alpha\,k}{r_+^2}\right]\,,
\ee where $A_H=V_{n-1}\,r_+^{n-1}$. This derivation is essentially
an application of Eq.~(\ref{Eaction}) and second law of black hole
thermodynamics. So, conceptually, it is fundamentally different
from the calculation in~\cite{Cai01a} where entropy comes from
first law. Eq.~(\ref{entropyGB}) is the correct entropy formula
even in a flat spacetime ($\Lambda=0$)~\cite{Myers88a}, so the
cosmological constant on the AdS boundary is not dynamical. As a
result, the central charge of an effective theory with a GB term
allows one to compute entropy without breaking Virasoro algebra
near the horizon~\cite{Cvitan02a}. The entropy flow \be d {\cal
S}= \frac{(n-1)A}{4G_{n+1}\,r_+}\left(1+\frac{2\alpha\,
k}{r_+^2}\right)\,, \ee is always positive, because $r_+^2+2\alpha
k\geq 0$ should hold for black hole interpretation~\cite{IPN02a},
and satisfies a generalized second law~\cite{Bekenstein73a}.
Moreover, since \bea T{\cal S}&=&
\frac{(n-1)V_{n-1}r_+^{n-4}}{16\pi G_{n+1}\,(n-3)}
\Bigg[(n-2)kr_+^2+(n-4)\,\alpha k^2\nn \\
&{}& +\,\frac{n\,r_+^4}{l^2}\Bigg]-\frac{V_{n-1} r_+^{n-1}}{16\pi
G_{n+1}}\,\frac{8\pi\,T}{n-3}\,, \eea one readily evaluates the
thermodynamic energy to be \bea \label{netenergy} E &=&  T{\cal
S}+F=M-\frac{(n-1)V_{n-1}} {16\pi G_{n+1}}\,\mu_{e}
=M -M_{e}\,,\nn \\
M&=& \frac{(n-1)\,V_{n-1}}{16\pi G_{n+1}} \left(k\,
r_+^{n-2}+\frac{r_+^n}{l^2} +\alpha k^2\,r_+^{n-4}\right)\,. \eea
For $k=1$, since $M_e=0$, one has $E=M$. For $k=-1$, since
$M_e<0$, $E\neq M$, in general. It is quite interesting that, for
$k=-1$, $E=0$ at $r_+=r_e$, and $E>0$ otherwise. Consider for
concreteness the $n=4$ case. Then, one has \be \label{5denergy}
E=\frac{3V_3}{16\pi G}\,\mu +\frac{3l^2\,V_3}{64\pi
G}\left(1-\frac{4\alpha}{l^2}\right)\,.\ee This energy is
vanishing at the extremal state, and also in Nariai limit
$\mu=\alpha-l^2/4$. As in de-Sitter case~\cite{Vijay01a}, the
Nariai solution is not the ground state in $n\neq 4$.

The black hole entropy~(\ref{entropyGB}) is always positive for
the curvature $k=0,\,1$. However, for $k=-1$, one has \be
\label{entropy5} {\cal S}=\frac{V_{3,k=-1}\,r_+^{3}}{4G_{5}}
\left(1-\frac{6\alpha}{r_+^2}\right)\,\Rightarrow {\cal S}_{e}=
\frac{V_3}{G_5}\,\frac{l^3}{2^{7/2}}
\left(1-\frac{12\alpha}{l^2}\right)\,. \ee Thus, in particular,
when one approaches a massless state at $\alpha=l^2/4$, the
extremal entropy becomes negative.
This is of course not an encouraging situation, because, as a
microscopic interpretation, the black hole entropy is the
logarithm of the number of (quantum) states and should be
positive. It is expected that additional higher order corrections,
like that $R^4$ terms, might cure this problem, so that a full
theory will yield only positive (extremal) entropy. One also notes
that, for the $\alpha=0$ case, the $k=-1$ extremal ground state
has positive entropy~\cite{Emparan99a} \be {\cal
S}_{e}=\frac{V_{n-1}}{G_{n+1}}\,\frac{l^{n-1}}{2^{(n+3)/2}}\,. \ee
These results further provide a hint that a massless extremal
state is simply not allowed as ground state.

\begin{figure}[ht]
\begin{center}
\epsfig{figure=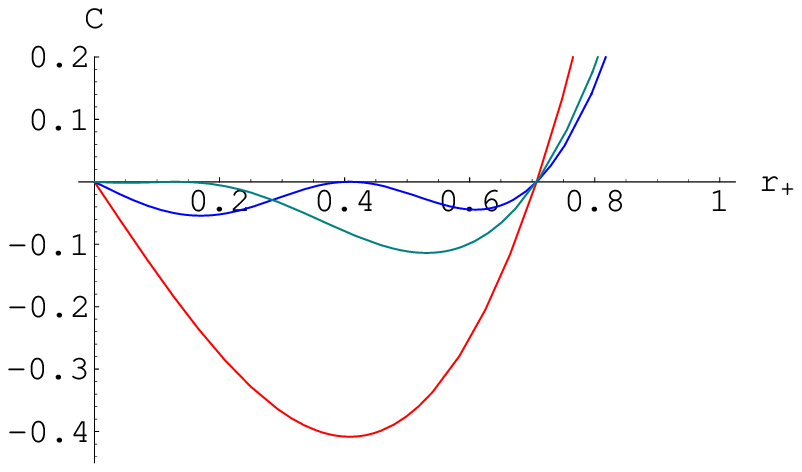, height = 4.1 cm, width = 7.0 cm}
\end{center}
\begin{center}
\epsfig{figure=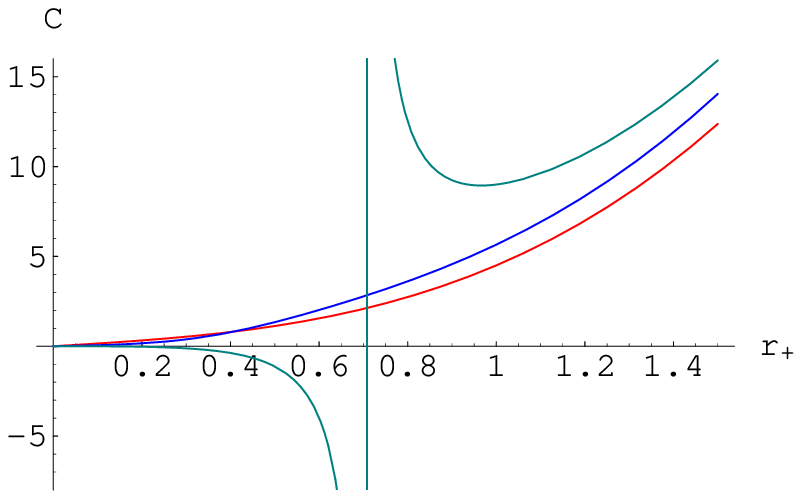, height = 4.1 cm, width = 7.0 cm}
\end{center}
\caption{The specific heat ($C=dE/dT$) {\it vs.} horizon radii.
The parameters are fixed as $l=1$, $V_{n-1}/4G_{n+1}=1$, $n=4$,
(a) $k=-1$ (upper plot): $\alpha=1/4$ (big single cusp),
$\alpha=1/12$ (two cusps), and $\alpha=1/120$ (small single cusp).
(b) $k=1$ (lower plot): the curve with $\alpha=0$ develops
singularity at $r_+=1/\sqrt{2}$, so a small (large) black hole has
negative (positive) specific heat, and two other curves correspond
to $\alpha=1/12$ and $\alpha=1/4$ (up to down).}
 \label{figure1}
\end{figure}

As the first plot in Fig.~(\ref{figure1}) shows, the small horizon
regime $r<r_{e}$ has a single branch for $\alpha=\frac{l^2}{4}$
and two branches for $\alpha<\frac{l^2}{4}$. The first branch
(cusp) on left, which might have negative specific heat, has no
black hole interpretation since this region is not allowed due to
a constraint $r_+^2>2\alpha$. Here we should note that, when
$k=-1$, $n=4$, for the coupling $\alpha=\frac{l^2}{12}$, the
Euclidean period $\beta$ is negative in the range
$\frac{1}{6}<r_+^2<\frac{1}{2}$, i.e., $0.408<r_+<0.707$. So the
Hawking temperature, which is a non-negative entity, should be
defined as $T=|\beta^{-1}|$. That is, in the range
$0.408<r_+<0.707$, the specific heat must be defined by
$C=-\beta^2\,dE/(d(-\beta))$. As a result, the second cusp in the
first plot of Fig.~(\ref{figure1}) should be mirror reflected, and
hence can have a positive specific heat. Nevertheless, for the
coupling $\alpha=\frac{l^2}{4}$, the Euclidean period $\beta$ is
always positive, so the formula $C=-\beta^2\,dE/(d\beta)$ is still
effective. For this particular coupling, the specific heat could
be negative, which might signal the instability of a massless
state. Because the energy condition $E\geq 0$ always holds, the
black holes of size of the extremal state or bigger than this have
zero or positive specific heat, and the corresponding solutions
are thermodynamically stable and globally preferred.

It is interesting that the minimum of the energy is also the
minimum of the temperature. As a result, the ratio $dE/dT$ is well
behaved even if $k=1$, which should be contrasted with the result
in Einstein gravity $(\alpha=0)$. This might show the emergence of
a stable branch of small spherical black holes, and similar result
was obtained by Caldarelli and Klemm in~\cite{Klemm99a}, where a
detailed treatment of M-theory/stringy-corrections, specifically,
the ${\cal O}({\alpha^\p}^3)$ corrections of type IIB string
theory, to black hole thermodynamics is presented. It has been
shown that~\cite{Klemm99a} the leading stringy or M-theory
corrections do not give rise to any phase transition for flat and
hyperbolic horizons, although to a quotient of hyperbolic space
$H^{n-1}/\Gamma$ there may arise new phase transitions. Further
elaboration and related discussion upon this issue appear
in~\cite{IPN03a}.

Given the importance of Gauss-Bonnet correction to Einstein
gravity, the extremal entropy is non-negative only if
$12\alpha<l^2$. This constraint also enforces the positivity of
energy for the $k= 1$ case. Following~\cite{Vijay01a,Nojiri01c},
we may calculate the total mass (quasi-local energy) of $k=+1$
Schwarzschild anti-de Sitter spacetime using the surface
energy-momentum tensor. In $n+1=5$, we find \be \label{5dnewE}
E=\frac{3 V_3\, l^2}{16\pi G}\left(1-\frac{12\alpha}{l^2}\right)
\left(\frac{1}{4}+\frac{\mu}{l^2}\right)\,.\ee In using a relation
$d{\cal S}=\beta\,dE$, we arrive at \be {\cal S}=\int
\frac{dr_+}{T}\, \frac{dE}{d\mu}\,\frac{d\mu}{dr_+}=\frac{V_3}{4G
l^2}
\left(r_+^3+6\alpha\,r_+\right)\left(l^2-12\alpha\right)+{\cal
S}_0\,. \ee This entropy is non-negative when $\alpha <l^2/12$,
since ${\cal S}_0=0$ at $r_+=0 $. It is worth noting that the
positivity of energy and entropy in the $k=+1$ case also ensures
that the $k=-1$ extremal entropy is non-negative.

Next we consider the action~(\ref{5dAdS}). The metric solution
that solves the field equations, to the order ${\cal
O}(\varepsilon)$, is \be f(r)= k-\frac{m}{r^2}+\frac{r^2}{l^2}
+\frac{\varepsilon\,m^2}{r^6}\,,\label{higher1}\ee where $l^2$ is
related to $\Lambda$ via $l^2\left(\Lambda\,l^2+6\right) =
2\,\varepsilon$. Using $f(r_+)=0$, we express $m$ as a function of
horizon radius: \be \label{bhmassm} m=
\left(k+\frac{r_+^2}{l^2}\right)\left[r_+^2+\varepsilon
\left(k+\frac{r_+^2}{l^2}\right)\right]\equiv \frac{4\pi^2
l^3}{N^2}\,\frac{M}{3V_3}\,, \ee where $r_+$ is the largest
positive root of $f(r)$ and $M$ is the black hole mass. The
inverse Hawking temperature is \be \label{higher5} \beta=
\frac{1}{T}=\frac{2\pi\,r_+^3\,l^4}{r_+^2\,l^2\left(k\, l^2+2
r_+^2\right) -2\varepsilon \left(k\, l^2+r_+^2\right)^2}\,. \ee
The extremal $k=-1$ solution, implied by $T=0$, reads \be
\label{Extremal} r_{e}^2\simeq \frac{l^2}{2}\,
\left(1+\frac{\varepsilon}{l^2}\right)\,,\quad m_{e}\simeq
-\frac{l^2}{4}\left(1-\frac{\varepsilon}{l^2}\right)\,. \ee Using
these as background values for $k=-1$, we obtain an Euclideanized
action valid for $k=0,~\pm 1$ to be~\cite{quote} \bea
\label{actionk-1} \hat{I} &=& \frac{V_3\,N^2} {4\pi^2\,
l^3}\,\Bigg[\left(m-\frac{2\,r_+^4}{l^2}\right)
\left(1-\frac{2\varepsilon}{l^2}\right)
-\frac{6\varepsilon\,m^2}{r_+^4}\nn \\
&{}& +\frac{3\,l^2}{4} \left(1+\frac{\varepsilon}{l^2}\right)
\Bigg] \,. \eea  The last expression, independent of $r_+$, will
be in effect only to the $k=-1$ case. As usual, free energy is
defined by $\hat{I}=\beta F$. The resulting entropy is \be
\label{entropyRiemann} {\cal S}= \beta^2\,\frac{\partial
F}{\partial\beta}= \frac{V_3\,r_+^3}{4}\,\frac{4 N^2}{\pi
l^3}\left[1+\frac{1}{4N}\, \frac{2r_+^2+3k\,l^2} {r_+^2}\right]\,.
\ee This entropy is essentially positive in the limit $r_+
>>l$. It may be negative for $k=-1$ when
$r_+^2 <\frac{3\,l^2}{2}$, but this limit is not allowed due to
the energy condition $E\geq 0$. In the large $N$ limit,
(\ref{entropyRiemann}) approximates to usual form ${\cal
S}=\frac{A_H}{4G}$. So one can expect that for large black holes
the asymptotic regions feel only minor corrections due to the
higher curvature terms. The extremal entropy \be {\cal S}_{e}
=\frac{V_3}{4G}\,\frac{l^3}{2\sqrt{2}}\left(1-\frac{1}{N}\right)
\ee is positive since $N >1$. The thermodynamic energy is \bea
\label{totalE} E &=&\frac{\partial \widehat{I}}{\partial
\beta}=F+T{\cal S}=M+E_{k}\,, \\
 E_{k}&=&\frac{3V_3\,N^2}{4\pi^2
l^3} \left[\frac{2\varepsilon\,r_+^4}{l^4}
\left(1+\frac{kl^2}{r_+^2}\right)+\frac{k^2\,l^2}{4}
\left(1+\frac{\varepsilon}{l^2}\right)\right]\,.\nn
\label{energyk} \eea One reads $M$ from~(\ref{bhmassm}). The
specific heat $C=\frac{\partial E}{\partial T}$ is \bea
C&=&\frac{3\,V_3\,r_+^3\,N^2}{\pi\,l^3}\Bigg[
\frac{2r_+^2\left(l^2+3\varepsilon\right)}{l^2\left(2r_+^2-k\,l^2\right)}
+\frac{k\left(l^2+4\varepsilon\right)}{2r_+^2-k\,l^2}\nn \\
&+& \frac{2\,\varepsilon\left(l^2+2k\,r_+^2\right)
\left(3k^2\,l^4+2k\,r_+^2 l^2-r_+^4\right)}
{r_+^2\,l^2\left(l^2-2k\,r_+^2\right)\left(2r_+^2-k\,l^2\right)}
\Bigg]\,. \eea A pleasing result is that the energy and specific
heat are vanishing at the extremal state defined by
(\ref{Extremal}), an important hint that the extremal state is the
ground state. For $k=1$, there is a discontinuity in specific heat
at $r_+=\frac{l}{\sqrt{2}}$, even if $\varepsilon>0$. This is
partly because the solutions are only perturbative and we have
retained the terms only linear in $\varepsilon$. In the $k=-1$
case, however, the solutions are well behaved, for example, the
specific heat and entropy are positive when $r_+>r_{e}$. A
difference from the $\varepsilon=0$ case is that now a small size
black hole has a positive specific heat at finite coupling
$3<N<\infty$.

\medskip
We end with few remarks and future problems.

We have calculated leading order curvature corrections to the
black hole entropy with horizons $k=0,\,\pm 1$. In general, the
entropy is not obtained by evaluating the horizon area of the
unperturbed solution divided by $4G$. It is encouraging that the
formulae (\ref{entropyGB}), (\ref{entropyRiemann}) perfectly match
with the entropies calculated using Wald's covariant
approach~\cite{Wald93a}, where the entropy is (unambiguously)
determined by a local geometric expression at the horizon.
Presumably, these results provide some elegant test of our
knowledge of entropy in string theory, for the higher curvature
terms as the Gauss-Bonnet invariant and/or
$R_{\mu\nu\lambda\rho}R^{\mu\nu\lambda\rho}$ interaction arise in
most string theory as leading $\alpha^\p$-corrections.

In general, the hyperbolic AdS black hole with zero (extremal)
mass is not stable as a supersymmetric background. The stability
of a hyperbolic horizon is therefore an important issue in
dimensions $n+1> 4$, which might be essential for a
non-supersymmetric extension of AdS/CFT correspondence. We find
that a negative mass $k=-1$ extremal background, which has the
lowest energy configuration in its asymptotic class, is stable
under gravitational perturbations when $\frac{\alpha}{l^2}<<1$,
and the potential is bounded from below (work in preparation). It
would be interesting in this case to investigate the thermal phase
structures and conformal behavior at infinity by coupling the
theory with scalars.

\medskip {\it Acknowledgements}:
I am grateful to Juan Maldacena and Soo-Jong Rey for providing
many intuitions on the problem; D. Birmingham and S. Nojiri for
fruitful comments on the subject. I wish to acknowledge a warm
hospitality of CERN Theory Group during my visit. This work is
supported by the National Science Council, the center for
Theoretical Physics at NTU, Taiwan, R.O.C..


\begin{thebibliography}{100}
\bibitem{quote} The Euclidean action derived
in~\cite{IPN02a} (see also~\cite{Nojiri01b}) is applicable only to
the $k=\{0,\,1\}$ horizons.
\end{thebibliography}
\end{document}